\documentclass{desyproc}

\begin{document}
\title{Exclusive meson pair production and the gluonic component of the eta, eta$^\prime$ mesons}

\author{{\slshape \footnote{Speaker} L.A. Harland--Lang$^{1}$, V.A. Khoze$^{1,2}$, M.G. Ryskin$^{2}$}\\[1ex]
$^1$Institute for Particle Physics Phenomenology, University of Durham, Durham, DH1 3LE\\
$^2$Petersburg Nuclear Physics Institute, NRC Kurchatov Institute, Gatchina, St. Petersburg, 188300, Russia  }

\contribID{Harland-Lang\_Lucian}


\acronym{EDS'09} 

\maketitle

\begin{abstract}
We present the results of recent work on the central exclusive production of meson pairs. We concentrate on the case of flavour--singlet pseudoscalar mesons $\eta$, $\eta'$ and show that the central exclusive process, as modelled using a recent novel application of the `hard exclusive' perturbative formalism, is potentially highly sensitive to the size of the gluon content of these states. We also discuss the exclusive production of meson pairs in the lower mass region, and present the first results of the new \texttt{Dime} Monte Carlo. 
\end{abstract}

\section{Introduction}

Central exclusive production (CEP) processes of the type
\begin{equation}\label{exc}
pp({\bar p}) \to p+X+p({\bar p})\;,
\end{equation}
can significantly extend the physics programme at hadron colliders. These reactions represent an experimentally very clean signal and provide a very promising way to investigate both QCD dynamics and new physics in hadron collisions. They have been widely discussed in the literature, and we refer the reader to~\cite{Martin:2009ku,Albrow:2010yb,HarlandLang:2013jf} for reviews and further references. 

A particularly interesting example, which has been the topic of recent investigations in~\cite{HarlandLang:2011qd,Harland-Lang:2013ncy,Harland-Lang:2013qia}, is the production of light meson pairs ($X=\pi\pi, KK, \rho\rho, \eta(')\eta(')$...) at sufficiently high transverse momentum $k_\perp$ that a perturbative approach can be taken. In Section~\ref{eta} we consider the case of $\eta(')\eta(')$ production, and show how this can be highly sensitive to the gluonic component of the $\eta$, $\eta'$ mesons. We recall that currently, while different determinations of the $\eta$--$\eta'$  mixing parameters are generally consistent, the long--standing issue concerning the extraction of the gluon content of the $\eta'$ (and $\eta$) remains uncertain, see~\cite{Harland-Lang:2013ncy} for more details and references.

More generally, theoretical studies of meson pair CEP within a `non--perturbative' (calculated in a Regge theory) framework in fact have a long history, see~\cite{HarlandLang:2012qz} and references therein for more details. Such an approach should be relevant at lower values of the meson transverse momentum $k_\perp$, where most of the data is expected to lie. We present in Section~\ref{dimesec} the first results from the new \texttt{Dime} Monte Carlo~\cite{dime} for meson pair CEP, modelled within such a Regge--based approach. We consider the case of $\pi^+\pi^-$ production and show how it may be used as a probe of the phenomenological production model. We also demonstrate how, in the presence of tagged protons, such processes may be used as a test of the models of hadronic interactions needed to calculate the soft survival factors.

\section{The gluonic component of the $\eta$, $\eta'$}\label{eta}

The $g(\lambda_1)g(\lambda_2)\to M_1 M_2$ helicity amplitudes relevant to the CEP of light meson pairs $M_1$, $M_2$ are calculated using an extension of the `hard exclusive' formalism, see e.g.~\cite{Brodsky:1981rp}. The basic idea is that the hadron--level amplitude can be written as a convolution of a perturbatively calculable parton--level amplitude, $T_{\lambda_1\lambda_2}$, for the $g(\lambda_1)g(\lambda_2)\to q\overline{q}(gg)\,q\overline{q}(gg)$ process (where each $q\overline{q}$ or $gg$ pair is collinear with the meson momentum and has the appropriate colour, spin, and flavour content projected out to form the parent meson), and a `distribution amplitude' $\phi(x)$, which contains all the non--perturbative information about the binding of the partons in the meson. 

 \begin{figure}
\begin{center}
\includegraphics[scale=0.5]{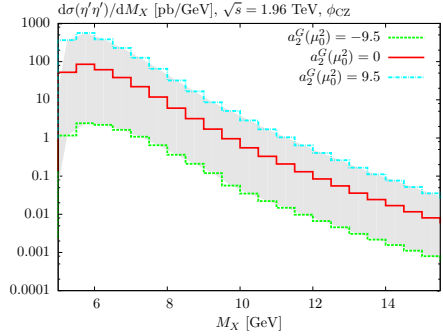}
\caption{Differential cross section ${\rm d}\sigma/{\rm d}M_{\eta'\eta'}$ for $\eta'\eta'$ production  at $\sqrt{s}=1.96$ TeV, taken from~\cite{Harland-Lang:2013ncy}.}\label{etamcz}
\end{center}
\end{figure}

As discussed in~\cite{Harland-Lang:2013ncy}, the flavour--singlet $\eta, \eta'$ mesons should contain a non--zero $gg$ component, which can be accessed via the  $gg\to gg\,q\overline{q}(gg)$ parton--level diagrams. The relevant amplitudes were calculated in~\cite{Harland-Lang:2013ncy}, giving
\begin{align}\label{lad0}
T_{++}^{qq}&=T_{--}^{qq}=\frac{\delta^{ab}}{N_C}\frac{64\pi^2 \alpha_S^2}{\hat{s}xy(1-x)(1-y)}\frac{(1+\cos^2 \theta)}{(1-\cos^2 \theta)^2}\;,\\ \label{tgq0}
 T^{gq}_{++}&=T^{gq}_{--}=2\,\sqrt{\frac{N_C^3}{N_C^2-1}}(2x-1)\cdot\,T_{++}^{qq}\;,\\ \label{tgg0}
T^{gg}_{++}&=T^{gg}_{--}=4\,\frac{N_C^3}{N_C^2-1}(2x-1)(2y-1)\cdot\,T_{++}^{qq}\;,
\end{align}
corresponding to the $gg \to q\overline{q}q\overline{q}$, $ggq\overline{q}$, $gggg$ final states, respectively. Remarkably, it was observed that the amplitudes, despite coming from completely separate classes of Feynman diagrams, are identical in form up to overall normalization factors. One important phenomenological consequence of this is that we expect no dynamical suppression in the $gg$ final states. This motivated a detailed a numerical study, performed in~\cite{Harland-Lang:2013ncy}, in which it was shown that the CEP cross sections for $\eta(')\eta(')$ states could indeed be highly sensitive to any $gg$ component. In particular, guided by the fit of~\cite{Kroll:2012hs},  a band of representative values for the size and sign of the $gg$ wavefunction $\phi_g(x)$ within the $\eta'$, $\eta$ were taken, and the predicted CEP cross sections were shown to vary by up to an order of magnitude, as shown in Fig.~\ref{etamcz}. We may therefore hope that future $\eta'$, $\eta$ pair CEP data and analysis will be 
forthcoming from the Tevatron and the LHC and that through this we can shed some light on this interesting and currently uncertain question.

\section{The `non--perturbative' regime}\label{dimesec}

As discussed in the Introduction, the predicted cross sections for meson pair CEP are in general much larger in the lower $k_\perp$ ($M_X$) region, for which the perturbative approach is not necessarily applicable. At this conference the first data on exclusive $\pi^+\pi^-$ production from CDF~\cite{cdfdat} and the STAR collaboration at RHIC, with tagged protons~\cite{STAR}, have been discussed. In this lower mass region we can apply a phenomenological model, in which the mesons are produced via double Pomeron exchange, and the cross section can be calculated using the tools of Regge theory, see~\cite{HarlandLang:2012qz}. The details of such a model are currently somewhat uncertain; in particular, prior to the CDF data, the form factor for the coupling of the Pomeron to the meson pair production subprocess was relatively unconstrained, in particular as the meson $k_\perp$ increases. 

\begin{figure}
\begin{center}
\includegraphics[scale=0.5]{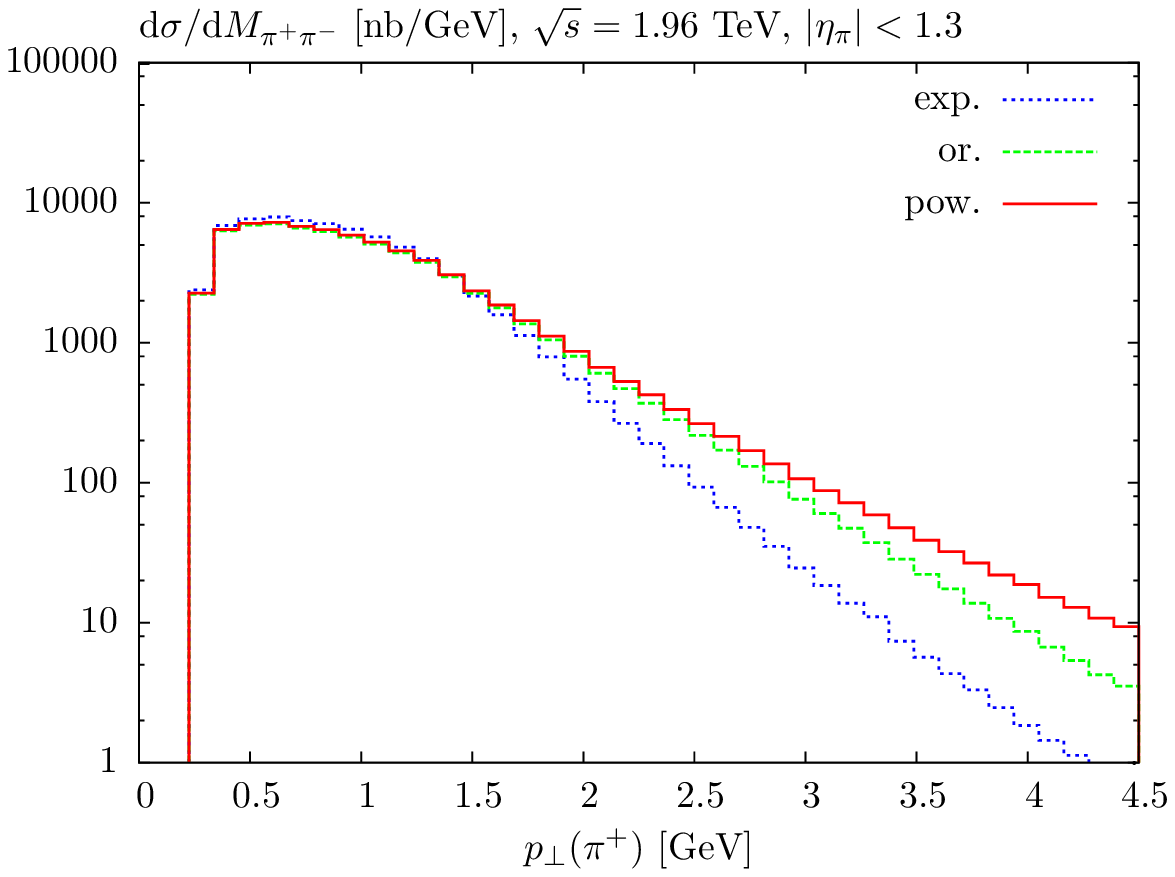}
\includegraphics[scale=0.5]{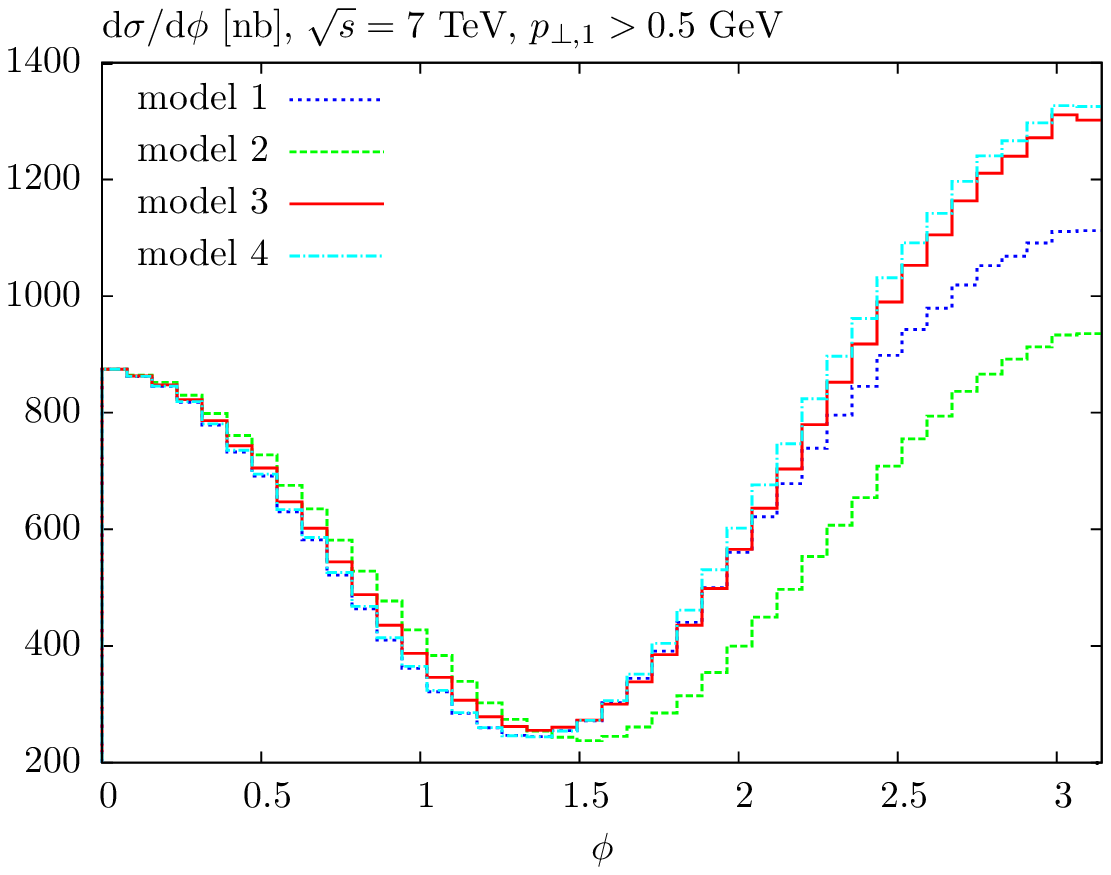}
\caption{Left: $M_{\pi\pi}$ distribution for $\pi^+\pi^-$ production at $\sqrt{s}=1.96$ TeV. Right: distribution in the relative azimuthal angle $\phi$ between the outgoing protons for $\pi^+\pi^-$ at $\sqrt{s}=7$ TeV, for different models choices for the soft survival factor, as defined in~\cite{Khoze:2013dha}.}\label{piplots}
\end{center}
\end{figure}

We have recently implemented such a model in the new \texttt{Dime} MC~\cite{dime}, including different possibilities for this form factor, and in Fig.~\ref{piplots} (left) we show the predicted invariant mass $M_{\pi\pi}$ distribution for $\pi^+\pi^-$ production at $\sqrt{s}=1.96$ TeV for these different choices, defined in~\cite{dime}, which we can compare with the CDF data to shed light on this model. We also show in Fig.~\ref{piplots} (right) the distribution, at $\sqrt{s}=7$ TeV, in the relative azimuthal angle $\phi$ between the outgoing protons, for different models choices for the soft survival factor, as defined in~\cite{Khoze:2013dha}. In this Figure, a cut of $p_\perp>0.5$ GeV has been placed on the transverse momenta of one proton, and in this region we can see an interesting `diffractive dip' structure, which is driven purely by the effect of the soft survival factor. Indeed, we can see that the precise shape of this is somewhat sensitive to the specific model choice. Thus, the observation of for example $\pi^+\pi^-$ CEP in the presence of tagged protons would provide novel insight into the models of soft physics used to calculate these survival factors.

LHL and VAK thank the conference organizers for support and for a very interesting and productive conference.

\begin{footnotesize}

\end{footnotesize}
\end{document}